\documentclass[a4paper,11pt]{article}
\usepackage{jheppub} 

\DeclareMathOperator{\Tr}{Tr}

\def\diag{\mathop{\rm diag}\nolimits}

\def\beq#1\eeq{\begin{align}#1\end{align}}

\title{$\mathcal{N}=1$ curve }

\author[1,2]{Dan Xie}

\affiliation[1]{Center of Mathematical Sciences and Applications, Harvard University, Cambridge, 02138, USA}
\affiliation[2]{Jefferson Physical Laboratory, Harvard University, Cambridge, MA 02138, USA}

\abstract{$\mathcal{N}=1$ curve is defined for four dimensional class  ${\cal S}$ theory 
 using  Cayley-Hamilton theorem for two commuting matrices.  The curve consists of three ingredients: 1: A set of N+1 degree N equations defining a curve; 2: a set of constraints relating the 
 coefficients in the curve; 3: a canonically defined differential. 
 We then extract from  spectral curve various physical information such as the space of moduli fields, chiral ring relations, full moduli space, etc. 
Many examples are discussed, and the curve recovers the intricate vacua structure which often involves highly non-trivial field theory dynamics such as
monopole condensation, dynamical generated superpotential,
Seiberg duality, etc. }

\begin{document} 
\maketitle
\flushbottom

\section{Introduction}
One of distinguished feature of supersymmetric field theories is that they have quantum moduli space of vacua. 
This kind of phenomenon is  especially interesting for  four dimensional  theory as  all kinds of interesting  phase structures can occur in the infrared \cite{Intriligator:1995au}. 
The structure of moduli space is crucial for understanding the dynamics of these theories. 

The  low energy effective theory on  Coulomb branch of $\mathcal{N}=2$ theory can be determined effectively by writing down a Seiberg-Witten (SW) curve \cite{Seiberg:1994rs,Seiberg:1994aj}. The most effective ways of finding a 
$\mathcal{N}=2$ curve are  using string theory tools \cite{Witten:1997sc,Katz:1997eq} and the connection to integrable system \cite{Donagi:1995cf} (those two approaches are closely related). Among these approaches, the Hitchin 
integrable system \cite{Hitchin:1987bc} plays an important role, and the SW curve derived from the spectral curve of the Hitchin system takes the following general form:
\begin{equation}
v^N+\sum_{i=2}^N \phi_i(z, u_j) v^{N-i}=0;
\label{N2}
\end{equation}
Here $u_j$ are the moduli fields parameterizing the Coulomb branch. There is also a canonical SW differential $\lambda=v dz$ defined on the curve. 
Once a SW curve is written down, one can determine the low energy effective actions from the complex structure of the curve, central charges of BPS particles using the SW differential, and various interesting 
physics at the singular points, etc. 

It was soon realized in \cite{Intriligator:1994sm} that one can write down a similar SW curve for Coulomb branch of  $\mathcal{N}=1$ theory. Later
on M5 brane construction is used to write down  curves for theory in confining and Higgs phase \cite{Witten:1997ep,Hori:1997ab}, and a $\mathcal{N}=1$ curve also 
appeared in the Dijkgraff-Vafa conjecture for some $\mathcal{N}=1$ theories \cite{Dijkgraaf:2002fc}. These methods  more or less
rely on $\mathcal{N}=2$ SW curve and it is hard to write down curves for more general  $\mathcal{N}=1$ theories. There are also no
integrable system tools available.

In \cite{Xie:2013rsa}, we proposed a method to write down  curves for more general $\mathcal{N}=1$ theory engineered using M5 brane without
relying on the knowledge of $\mathcal{N}=2$ theory. The new feature is that the moduli fields satisfy intricate relations, which is 
probably the reason why $\mathcal{N}=1$ curve is much harder to find. There is no such issue for $\mathcal{N}=2$ Coulomb branch as there are 
no relations between the moduli fields. 
So the task of writing down $\mathcal{N}=1$ curve consists of two steps: first write down a family of  curves, and 
second find the relations between the coefficients which encode the moduli fields. The difficult part is the second step, which is only solved for rank one theory in \cite{Xie:2013rsa}.
The purpose of this paper is to find the relations for general rank N theory.

$\mathcal{N}=1$ class $\mathcal{S}$ theory is engineered by wrapping N M5 branes on a Riemann surface $\Sigma$ which is embedded in a local Calabi-Yau
three-fold. The normal deformation of the Riemann surface is encoded by a rank two holomorphic vector bundle ${\cal N}$ \footnote{In this paper, we assume that the rank two bundles 
are split, i.e. ${\cal N}=L_1\bigoplus L_2$. A detailed discussion about the choices of these bundles and their physical meaning will appear in \cite{xie2014A}.} whose determinant is equal to the canonical bundle of 
Riemann surface, therefore the infrared deformations are encoded by two Higgs fields $\Phi_1, \Phi_2$. The most crucial fact of these Higgs fields is 
that they have to commute with each other $ [\Phi_1, \Phi_2]=0$  \cite{Xie:2013gma}. 

The infrared physics is encoded by the spectral curve of $\Phi_1$ and $\Phi_2$.  Let's use $z$ to denote the coordinate on $\Sigma$ and $v, w$ to denote the fibre coordinates of the
rank two bundle. The spectral curve encoding the eigenvalues of $\Phi_1$ and $\Phi_2$ can be thought of as another holomorphic curve $S$ embedded in the local Calabi-Yau 
parameterized by $v,w,z$.  

The spectral curve (\ref{N2}) of $\mathcal{N}=2$ theory is written down using the characteristic polynomial of a single Higgs field $\Phi_1$, and  $\Phi_1$ itself satisfies the characteristic polynomial equation due to Cayley-Hamilton 
theorem. This motivates us to 
find similar matrix equations for $\Phi_1, \Phi_2$ and then replace the matrices in the equations by the coordinates $v$ and $w$ to define 
$S$. The problem of finding such matrix equations is actually solved in \cite{procesi1976invariant,razmyslov1974trace}, 
and using their results $S$  is defined by following $N+1$ equations:
\begin{align}
v^aw^b+\sum_{i+j=2}^N c_{i,j}^{a,b} f_{ij}(z, u)v^{a-i}w^{b-j}=0;~~~~a+b=N; ~~a,b\geq0;
\label{cur}
\end{align}
Here coefficients $f_{ij}$ are holomorphic sections of line bundle $L_1^i \bigotimes L_2^j$, $u$ are the moduli fields,  and  $c_{i,j}^{a,b}$ are fixed constants.  
These equations can be derived using the generalized Cayley-Hamilton theorem for two commuting matrices, 
and the coefficients are  expressed in terms of 
the trace of  $\Phi_1$ and $\Phi_2$.  

These coefficients $f_{i,j}$ are not independent. Since they are expressed in terms of traces of $\Phi_1$ and $\Phi_2$, the constraints can be expressed in terms of the 
trace identities which can be found using the matrix equations, and  the full set of constraints take the following form: 
\begin{align}
& \sum_d g_d f_{a_1b_1}f_{a_2b_2}\ldots f_{a_rb_r}=0; ~~~~~  \sum a_i= m,~~\sum b_i=n,\nonumber\\
& N+2 \leq m+n\leq 2N;~~m\geq 2~or~n \geq 2.
\label{con}
\end{align}
Here the sum is over all possible partitions and $g_d$ are constants (could be zero). Each equation is labeled by  $(m,n)_i$ with $m,n$ indicating 
that the equation is a section of $L_1^m \bigotimes L_2^n$, and $i$ indicates different type of equations for fixed $(m,n)$. 
The equation \ref{cur} and the constraints \ref{con} are what we called $\mathcal{N}=1$ curve. The explicit curves and constraints of $A_1, A_2, A_3$ theories 
are presented in [\ref{curA1}, \ref{conA1},  \ref{curA2}, \ref{conA2}, \ref{curA3}, \ref{conA3}]. The interesting physics of $\mathcal{N}=1$ theory can be extracted simply by solving the constraints
due to the following crucial facts: the coefficients $f_{ij}$ are holomorphic sections!

There is also a canonically defined differential on the curve
\begin{equation}
\Omega=dv\wedge dw \wedge dz, 
\end{equation}
which is nothing but the holomorphic $(3,0)$ form on local Calabi-Yau three fold. This differential 
 can be used to calculate the effective superpotentials, domain wall tension, and  scaling dimension, etc.
 
We apply our method to many examples including $\mathcal{N}=2$ theory deformed by various superpotentials, and our curves give
exactly the same results derived from other methods. Those results are highly non-trivial, and often involve interesting dynamics including 
monopole condensation, dynamical generated superpotential, Seiberg duality, etc. 

This paper is organized as follows: in section 2, we present our method of writing down the spectral curve and the constraints. In section 3, 
we discuss how to find the full moduli space of $\mathcal{N}=1$ theory. Section 4 describes how to find the spectral curve for theory with 
various superpotentials. Finally, a summary and possible future direction is given in section 5.

\newpage
\section{Solving $\mathcal{N}=1$ curve}

\subsection{6d construction and spectral curve}
Four dimensional $\mathcal{N}=1$ class ${\cal S}$ theory is defined by compactifying 6d $(2,0)$ theory on a Riemann surface with 
various regular and irregular defects.  The data defining the theory can be summarized as follows:
\begin{itemize}
\item Choose 6d (2,0) theory of ADE type  and compactify it on a Riemann surface $\Sigma$ with $n$ punctures.
\item Choose regular and irregular defects at the puncture, and choose a pair of integers
($n_1$, $n_2$) to  represent  the number of defects of type I and type II \footnote{ Type I (Type II) means that Higgs field $\Phi_1$ ($\Phi_2$) is singular, and they can be singular at the same point.}.  We  only use local $\mathcal{N}=2$ punctures in this section.
\item Choose a pair of holomorphic line bundles $L_1$ and $L_2$ \footnote{In general, one can choose an arbitrary rank two holomorphic bundles, but the split case plays 
a special role, see \cite{xie2014A}} so that $\mathcal{N}=1$  (Calabi-Yau) condition is satisfied:
\begin{equation}
L_1\bigotimes L_2=K\bigotimes {\cal O}(\sum_{i=1}^n p_i);
\end{equation}
Here $K$ is the canonical bundle of the punctured Riemann surface. $\Phi_1$ and $\Phi_2$ are sections of following bundles:
\begin{equation}
L_1=L_1^{'}\bigotimes {\cal O}(\sum_{i=1}^{n_1} q_i),~~~~L_2=L_2^{'}\bigotimes {\cal O}(\sum_{i=1}^{n_2} r_i).
\end{equation}
Here $q_i, r_i$ are subsets of $p_i$. The conditions on the degrees are then 
\begin{equation}
deg(L_1^{'})+deg(L_2^{'})+n_1+n_2=2g-2+n.
\end{equation}
If $n_i\neq 0$, we require that the degree of $L_i$ to be non-negative, so it admits a meromorphic section.
\end{itemize}
The $\mathcal{N}=1$  theory space constructed in this way is remarkably large, and more details about these new theories will appear in \cite{xie2014A,xie2014B}, see also 
\cite{Benini:2009mz, Tachikawa:2011ea, Bah:2011je,Bah:2011vv,Bah:2012dg,Beem:2012yn,Gadde:2013fma,Maruyoshi:2013hja,Bonelli:2013pva, Bah:2013aha,Yonekura:2013mya,Agarwal:2013uga, Agarwal:2014rua, Giacomelli:2014rna,McGrane:2014pma} for some recent studies on
a small sample of these theories. 

We are interested in the moduli space of vacua  of those theories, which is conjectured to be governed by the following generalized Hitchin equation \cite{Xie:2013gma}:
\begin{align}
&F_{z\bar{z}}+h_1[\Phi_1, \Phi_1^{*}]+h_2[\Phi_2, \Phi_2^{*}]=0; \nonumber\\
&D_{\bar{z}} \Phi_1=D_{\bar{z}} \Phi_2=0; \nonumber\\
&[\Phi_1,\Phi_2]=0;
\end{align}
Here  $A_{\bar{z}}$ defines a rank $N$ holomorphic vector bundle, and $\Phi_i \in H^0(\Sigma, 
End(E)\bigotimes L_i)$, $i=1,2$ are 
two Higgs fields. $h_i$ are the proper Hermitian metric to make the first equation to be covariant. It is conjectured that the moduli space ${\cal M}_{GH}$ of above generalized Hitchin 
equation is related to the moduli space of four dimensional $\mathcal{N}=1$ theory: one can write down a spectral curve which can be identified with the 
$\mathcal{N}=1$ SW curve. This idea is the same as the one proposed in \cite{Witten:1997ep}:
 the infrared physics is determined by a different Riemann surface on which a single M5 brane wraps, which is the 
Seiberg-Witten curve; In fact, one has a family of  SW curve parameterized by the moduli fields. Mathematically, this means that the moduli space $M_{GH}$ has the following fibration structure
\begin{equation}
\pi:~{\cal M}_{GH}\rightarrow B;
\end{equation}
Here the coordinates on B is parameterized by the moduli fields. This fibration structure is nicely encoded by a spectral curve. 
Let's use $z$ to denote the coordinate of Riemann surface $\Sigma$, and $(v, w)$ to denote the tautological sections of two line bundles. Then the spectral curve 
is  a Riemann surface embedded in the total space of  rank two bundles parameterized by $(z, v, w)$, therefore the task is to find a set of polynomial 
equations depending on these three coordinates.  
We leave the full mathematical discussion about the moduli space of generalized Hitchin equation and the spectral curve to a different place \cite{xie2014C}, and here 
we take a more elementary way. 

Let's start with  $\mathcal{N}=2$ case where the spectral curve is known in full detail \cite{Hitchin:1987bc,Gaiotto:2009we}. In this case, we have $L_1=K, L_2={\cal O}$, and
 there is only one nontrivial Higgs field $\Phi_1 \in H^0(\Sigma, 
End(E)\bigotimes K)$ for the Coulomb branch. The spectral curve is nothing but the characteristic polynomial of $\Phi_1$. 
\begin{equation}
p(x)=\text{det}(x-\Phi_1)=x^N+\sum_{i=2}^N \phi_i x^{N-i}=0;
\end{equation}
here $\phi_i \in H^0(\Sigma, K^i)$ which can be expressed in terms of trace of matrix $\Phi_1$. 
There is a famous Cayley-Hamilton theorem which says that the matrix $\Phi$ satisfy the equations: $p(\Phi_1)=0$, and 
this fact is valid for any matrix!  We could find the spectral curve by first deriving a matrix equation for $\Phi_1$, and then 
replace $\Phi_1$ in the matrix equation by the coordinate $x$. 

This motivates the following strategy for finding the spectral curve for two $N\times N$ matrices: first find 
the matrix equations for  $\Phi_1$ and $\Phi_2$ in which the coefficients are expressed in terms of traces of $\Phi_1, \Phi_2$, 
and the spectral curve is found by replacing $\Phi_1$ and $\Phi_2$ with coordinates $v, w$. 
Fortunately, this problem has been solved by Procesi  and Razmyslov \cite{procesi1976invariant,razmyslov1974trace}, and they 
show that the generator for the matrix equations are actually finite! 

Even more importantly, using these matrix equations, one can find the trace identities for $\Phi_1$ and $\Phi_2$, and these 
trace identities will give us equations relating the coefficients in the spectral curve! The commuting condition on $\Phi_1$ and $\Phi_2$ plays an important role. 
In the following, we are going to discuss more details about the explicit construction of spectral curve.

$\textbf {A}_1$ \textbf{theory}:  
By $A_1$ theory we mean the four dimensional theories engineered using 6d $A_1$ $(2,0)$ theory.  
Let's start with the characteristic polynomial $p(\lambda)$ of a traceless $2\times 2$ matrix $X$:
\begin{equation}
p(\lambda)=\text{det}(\lambda-X)=0\rightarrow \lambda^2-{1\over 2} \Tr[X^2]=0.
\end{equation}
The Cayley-Hamilton theorem states that the matrix $X$ itself satisfies its characteristic polynomial equation:
\begin{equation}
p(X)=X^2-{1\over 2} \Tr[X^2]=0.
\end{equation}
If we have a pair of matrices $X_1, X_2$, then we have an extra matrix equation:
\begin{equation}
p(X_1+X_2)-p(X_1)-p(X_2)=0\rightarrow X_1X_2+X_2X_1-\Tr(X_1 X_2)=0;
\end{equation}
Let's apply the above matrix equations to two commuting matrices $\Phi_1$ and $\Phi_2$, and we have:
\begin{align}
& \Phi_1^2-{1\over 2}\Tr(\Phi_1^2)=0, \nonumber\\
& \Phi_1 \Phi_2-{1\over 2}\Tr(\Phi_1 \Phi_2)=0, \nonumber\\
& \Phi_2^2-{1\over 2}\Tr(\Phi_2^2)=0.
\end{align}
Notice that each term in the matrix equation has the fixed number of $\Phi_1$ and $\Phi_2$ factors, and we can label them by a pair of integers $(a,b)$, i.e. the first equation has 
a label $(2,0)$; and we also define the order of matrix equation as $a+b$. 
In fact, all the matrix equations are generated by the order 2 matrix equations, and the above three equations are enough for our purpose. 
From these equations, we can easily find a trace identity:
\begin{equation}
(\Tr\Phi_1 \Phi_2)^2=\Tr\Phi_1^2\Tr\Phi_2^2.
\label{tr1}
\end{equation}
Notice that the commuting condition plays an important role for 
deriving this equation. Moreover, this trace identity generates all the other trace identities. Once we find the full set of matrix equations, we can define the spectral curve by replacing $\Phi_1, \Phi_2$ with the coordinates $v, w$:
\begin{align}
& v^2=f(z), \nonumber\\
& v w= h(z), \nonumber\\
& w^2=g(z),
\label{curA1}
\end{align}
with 
\begin{equation}
f(z)={1\over 2}\Tr\Phi_1^2,~~~h(z)={1\over 2}\Tr\Phi_1 \Phi_2,~~~g(z)={1\over 2}\Tr\Phi_2^2.
\end{equation}
and we have the following constraints on the coefficients from the trace identity (\ref{tr1}):
\begin{equation}
h(z)^2=f(z) g(z).
\label{conA1}
\end{equation}
The spectral curve ($\ref{curA1}$) and the constraint ($\ref{conA1}$) are exactly what is found in \cite{Xie:2013rsa}. 

$\textbf {A}_2$ \textbf{theory}: 
Let's now consider $A_2$ theory and $3\times 3$ matrices. The characteristic polynomial for a traceless $3\times 3$ matrix $X$ is 
\begin{equation}
p(\lambda)=\lambda^3-{1\over2}\text{Tr}(X^2) \lambda-{1\over 3} \text{Tr}(X^3)=0.
\end{equation}
Again, the Cayley-Hamilton theorem says that $p(X)=0$. Now we use the following equations
\begin{equation}
p(X_1+X_2)+p(X_1-X_2)-2p(X_1)=0,~~~~p(X_1+X_2)-p(X_1-X_2)+2p(X_2)=0,
\end{equation}
to get another two matrix equations. Combine with the characteristic polynomials of $\Phi_1$ and $\Phi_2$, we get the following 
four order three matrix equations:
\begin{align}
&\text{$\Phi_1$}{}^{3}-\frac{1}{2} \text{$\Phi_1$} \text{Tr}[\text{$\Phi_1$}{}^{2}]-\frac{1}{3} \text{Tr}[\text{$\Phi_1$}{}^{3}]=0, \nonumber\\
&\text{$\Phi_1$}{}^{2}\text{$\Phi_2$}-\frac{1}{3} \text{$\Phi_1$}\text{Tr}[\text{$\Phi_1$}\text{$\Phi_2$}]-\frac{1}{6}\text{$\Phi_2$}\text{Tr}[\text{$\Phi_1$}{}^{2}]-\frac{1}{3}\text{Tr}[\text{$\Phi_1$}{}^{2}\text{$\Phi_2$}] =0,\nonumber\\
&\text{$\Phi_2$}{}^{2}\text{$\Phi_1$}-\frac{1}{3}\text{$\Phi_2$}\text{Tr}[\text{$\Phi_1$}\text{$\Phi_2$}]-\frac{1}{6}\text{$\Phi_1$}\text{Tr}[\text{$\Phi_2$}{}^{2}]-\frac{1}{3}\text{Tr}[\text{$\Phi_2$}{}^{2}\text{$\Phi_1$}] =0,\nonumber\\
&\text{$\Phi_2$}{}^{3}-\frac{1}{2} \text{$\Phi_2$} \text{Tr}[\text{$\Phi_2$}{}^{2}]-\frac{1}{3} \text{Tr}[\text{$\Phi_2$}{}^{3}]=0.
\end{align}
We have used the commuting condition on $\Phi_1$ and $\Phi_2$.  They can be  labeled as $(3,0), (2,1),(1,2), (0,3)$ based on the number of $\Phi_1,\Phi_2$ factors. Apparently, the above equations are 
the full set of order three matrix equations.  Using the matrix equation, our spectral curve is 
\begin{align}
\left(\begin{array}{c}
v^3\\
v^2w\\
vw^2 \\
w^3
\end{array}\right )+\left(\begin{array}{c}
f_{1}(z) v\\
f_{3}(z) v+{1\over3}f_1 w\\
f_{3}(z) w+{1\over 3}f_2 v\\
f_{2}(z) w
\end{array}\right )+\left(\begin{array}{c}
g_{1}(z)\\
g_{3}(z)\\
g_{4}(z)\\
g_{2}(z)
\end{array}\right )=0;
\label{curA2}
\end{align}
and the coefficients are expressed as the traces of two matrices:
\begin{align}
& f_1=-{1\over 2} \Tr \Phi_1^2,~~~~~g_1=-{1\over 3} \Tr \Phi_1^3,   \nonumber\\
& f_3=-{1\over 3} \Tr \Phi_1\Phi_2,~~g_3= -{1\over 3} \Tr\Phi_1^2\Phi_2,~~ g_4= -{1\over 3} \Tr\Phi_1\Phi_2^2, \nonumber\\
&f_2=-{1\over 2} \Tr \Phi_2^2,~~~~~~g_2=-{1\over 3} \Tr \Phi_2^3.
\end{align}
These traces are the full set of invariants of two commuting matrices. By invariant we mean the expressions which are invariant under the transformation
\begin{equation}
\Phi_1\rightarrow g \Phi_1 g^{-1},~~~~~\Phi_2\rightarrow g \Phi_2 g^{-1},
\end{equation}
with $g$ an arbitrary $3\times 3$ matrix. 
All the invariants are generated by the single trace and multiple trace of matrices \cite{procesi1976invariant,razmyslov1974trace}. Using matrix equations, we can express the invariants  such as $\Tr[\Phi_1^a\Phi_2^b], a+b>3$ 
 in terms of above generators. 

By multiplying the appropriate powers of $\Phi_1$ and $\Phi_2$ to one of matrix equation, then using other equation to eliminate the higher order term, finally, take the trace, we can get trace identity
 \footnote{For example, we can multiple $\Phi_1\Phi_2^2$ to the second equation, and then use 
the first and last equation to eliminate $\Phi_1^3$ and $\Phi_2^3$ term, we get a matrix equation involving $\text{Tr}(\Phi_1^3)\text{Tr}(\Phi_2^3)$; finally, we take the trace of this equation, and get
the $(3,3)$ trace identity.}.  Using this procedure, we find the following trace identities: 
\begin{align}
&\frac{1}{9}\text{Tr}\left[ \Phi_1^2\right] \text{Tr}\left[\Phi_2^2\right] \text{Tr}[\Phi_1\Phi_2]-\frac{1}{9} \text{Tr}[[\Phi_1\Phi_2]^3-\frac{1}{3} \text{Tr}\left[\Phi_1 \Phi_2^2\right] \text{Tr}\left[\Phi_1^2 \Phi_2\right]+
\frac{1}{3} \text{Tr}\left[\Phi_1^3\right] \text{Tr}\left[\Phi_2^3\right] =0,\nonumber\\
&-\frac{2}{3} \text{Tr}[\Phi_1 \Phi_2] \text{Tr}\left[\Phi_1^2 \Phi_2\right]+\frac{1}{3} \text{Tr}\left[\Phi_1^2\right] \text{Tr}\left[\Phi_1 \Phi_2^2\right]+\frac{1}{3} \text{Tr}\left[\Phi_1^3\right] \text{Tr}\left[\Phi_2^2\right]=0, \nonumber\\
&-\frac{2}{3} \text{Tr}[\Phi_1 \Phi_2] \text{Tr}\left[\Phi_1 \Phi_2^2\right]+\frac{1}{3} \text{Tr}\left[\Phi_1^2 \Phi_2\right] \text{Tr}\left[\Phi_2^2\right]+\frac{1}{3} \text{Tr}\left[\Phi_1^2\right] \text{Tr}\left[\Phi_2^3\right]=0, \nonumber\\
&\frac{1}{18} \text{Tr}[\Phi_1^2] \text{Tr}[\Phi_1 \Phi_2]^2 + \frac{1}{3} \text{Tr}[\Phi_1^2 \Phi_2]^2 - \frac{1}{18} \text{Tr}[\Phi_1^2]^2 \text{Tr}[\Phi_2^2] - 
\frac{1}{3} \text{Tr}[\Phi_1^3] \text{Tr}[\Phi_1 \Phi_2^2]=0, \nonumber\\
&
\frac{1}{18}  \text{Tr}[\Phi_2^2] \text{Tr}[\Phi_2 \Phi_1]^2 +  \frac{1}{3}  \text{Tr}[\Phi_2^2 \Phi_1]^2 -\frac{1}{18}  \text{Tr}[ \Phi_1^2] \text{Tr}[\Phi_2^2]^2 -  \frac{1}{3} \text{Tr}[\Phi_2^3] \text{Tr}[\Phi_2 \Phi_1^2]  =0.
\end{align}
Notice that each term also contains equal number of $\Phi_1$ and $\Phi_2$ factors for each equation, so we could label the above trace identity as type $(3,3), (3,2), (2,3), (4,2),(2,4)$. 
These are the full set of trace identities. 

Using the trace identities, we get the following constraints for the spectral curve of $A_2$ theory: 
\begin{align}
&\text{Type}[3][3]:~~- {4\over 9}f_1 f_2 f_3+f_3^3-g_3 g_4+g_1 g_2=0, \nonumber\\
&\text{Type}[3][2]:~~  -3f_3 g_3+f_1g_4+g_1f_2=0,\nonumber\\
&\text{Type}[2][3]:~~-3f_3g_4+f_2g_3+f_1 g_2=0,  \nonumber\\
&\text{Type}[4][2]:~~ f_1 f_3^2 -3g_3^2-{4\over 9} f_1^2 f_2   + 3 g_1 g_4=0, \nonumber\\
& \text{Type}[2][4]:~~ f_2 f_3^2 -3g_4^2-{4\over 9} f_2^2 f_1   + 3 g_2 g_3=0.
\label{conA2}
\end{align}

$\textbf {A}_{N-1}$ \textbf{theory}: 
In general, the spectral curve takes the following form:
\begin{align}
v^aw^b+\sum_{i+j=2}^N c_{i,j}^{a,b}  f_{ij}v^{a-i}w^{b-j}=0;~~~~a+b=N;
\label{curN}
\end{align}
Here $i, j$ is constrained such that the power of  $v, w$ is always non-negative, and $f_{ij}^{(a,b)}  \in H^0(\Sigma, L_1^iL_2^j)$, and  
$c_{i,j}^{a,b}$ are fixed constants. There are a total of $N+1$ equations. 

The above equations are  the consequence of the generalized Cayley-Hamilton theorem for two matrices due to Procesi and Razmyslov \cite{procesi1976invariant,razmyslov1974trace}, which claim that all the matrix equations of two $N\times N$ matrices  are generated by the following list:
\begin{equation}
\Phi_1^a\Phi_2^b+\sum_d c_d \Phi_1^{a_0}\Phi_2^{b_0} \Tr(\Phi_1^{a_1}\Phi_2^{b_1})\Tr(\Phi_1^{a_2}\Phi_2^{b_2})\ldots \Tr(\Phi_1^{a_r}\Phi_2^{b_r})=0,~~~~a+b=N;
\end{equation}
Here the sum is over all the partitions such that $\sum_{i=0}^ra_i=a,~~\sum_{i=1}^rb_i=b$, and $c_d$ are rational numbers which can be easily found. Using the above equation, 
$f_{ij}$ can be expressed in terms of trace of matrices:
\begin{equation}
f_{ij}=c\Tr[\Phi_1^i\Phi_2^j]+\ldots
\end{equation}
here $c$ is a fixed constant. By multiplying the appropriate power on one of above equation, then use other matrix equation to eliminate the higher order term, and finally  take the trace, we get the following trace identities:
\begin{equation}
\sum_d e_d \Tr[\Phi_1^{a_1}\Phi_2^{b_1}]\Tr[\Phi_1^{a_2}\Phi_2^{b_2}] \ldots \Tr[\Phi_1^{a_r}\Phi_2^{b_r}]=0;
\end{equation}
Here the power in each individual trace satisfies $a_i+b_i\leq N$, and $e_d$ are constants. 
The above equations can be expressed using the coefficients of the spectral curve, and the generators have the following form:
\begin{align}
& \sum_d g_d f_{a_1b_1}f_{a_2b_2}\ldots f_{a_rb_r}=0; ~~~~~ \sum a_i= m,~~\sum b_i=n,\nonumber\\
& N+2 \leq m+n\leq 2N;~~m\geq 2~or~n \geq 2.
\label{conN}
\end{align}
Here the sum is over all the partitions such that $\sum a_i=m,~\sum b_i=n$, and $g_d$ are some rational numbers (could be zero).  Each equation is labeled by  $(m,n)_i$ with $m,n$ indicating 
that the equation is a section of $L_1^m L_2^n$, and $i$ indicates different types of equations for fixed $(m,n)$. We have not seen this $i$ index for $A_1$ and $A_2$ theory, but 
they generically would appear, see an example in appendix.

There is a canonical differential defined on the spectral curve. Since coordinates $(z, v, w)$ parameterizes a local Calabi-Yau, it has a nowhere 
vanishing $(3,0)$ form: 
\begin{equation}
\Omega=dv\wedge dw \wedge dz;
\label{dif}
\end{equation}
and this differential is important for calculating the superpotential \cite{Witten:1997ep}, domain wall tension, scaling dimension \cite{Giacomelli:2014rna}, etc. The application of 
this differential will be discussed elsewhere. 

The spectral curve presented in (\ref{curN}), the constraints (\ref{conN}) of the coefficients, and the canonical differential (\ref{dif}) are 
the main result of this paper. The explicit equations for $A_3$ theory are presented in appendix  following above general procedure.

\subsection{Properties of spectral curve}
\subsubsection{Moduli fields and chiral ring relation}
The spectral curve presented in last subsection suggests that the moduli fields are given by the space of sections:
\begin{equation}
\bigoplus_{i,j}H^0(\Sigma, L_1^i L_2^j).
\end{equation}
The dimension of  holomorphic sections of a line bundle $L$ can be found using the Riemann-Roch theorem:
\begin{equation}
\dim H^0(\Sigma, L)-\dim H^0(\Sigma, K\bigotimes L^{-1})=\deg(L)-g+1;
\end{equation}
An important fact is that the the dimension of
 holomorphic sections of $L$ is zero if $deg(L)<0$. Using this fact and the Riemann-Roch theorem, we have the following simple results:
\begin{align}
&\dim H^0(\Sigma, L)=0,~~~~deg(L)<0,  \nonumber\\
&\dim H^0(\Sigma, L)=\deg(L)-g+1,~~~~deg(L)>2g-2,     \nonumber\\
&\dim H^0(\Sigma, L)=1,~~L={\cal O},    \nonumber\\
&\dim H^0(\Sigma, L)=0,~~deg(L)=0,~~L\neq {\cal O}. \nonumber\\
&\dim H^0(\Sigma, K)=g,~~~~L=K,  \nonumber\\
&\dim H^0(\Sigma, L)=g-1,~~~~deg(L)=2g-2,~L\neq K; 
\end{align}
Here ${\cal O}$ is the trivial bundle and $K$ is the canonical bundle. 
If $0<deg(L)< 2g-2$, the dimension of the holomorphic section depends on the specific line bundles. 

In many cases,  the spectral curve for a single Higgs field is important, and we can discuss more details. 
Consider spectral curve of a single Higgs field $\Phi\in H^0(\Sigma, End(E)\bigotimes L)$:
\begin{equation}
v^N+\sum_{i=2}^N\phi_i v^{N-i}=0;
\end{equation}
Let's denote the contribution of the $i$th puncture to the dimension of the moduli space as  $p_i$, then the dimension of the base is (assuming $deg(L)\geq2g-2$):
\begin{equation}
d_b=\sum_{punctures}p_i + deg(L){(N-1) (N+2)\over 2}-(N-1) (g-1).
\label{base}
\end{equation}

The next question is  what is the physical meaning of these moduli fields? In the $\mathcal{N}=2$ case, the moduli fields are the Coulomb branch operators, which 
are in particular invariant under the flavor symmetry. In the $\mathcal{N}=1$ case,  based on the analysis performed in \cite{deBoer:1997zy}, 
the moduli fields in the spectral curve describes the \textbf{gauge and flavor invariant} parts of the moduli space! 

The fields in $\bigoplus_{i,j}H^0(\Sigma, L_1^i L_2^j)$ are not independent, and they have to satisfy relations presented in last subsection, this leads to 
the chiral ring relations between the moduli fields. Many examples will be discussed in later sections.

\subsubsection{Genus of spectral curve and phases}
The genus of the spectral curve $S$ tells us how many massless photons are left in the low energy theory. 
The spectral curve can be thought of as a N-fold cover over the Riemann surface parameterized by $z$:
\begin{equation} 
S \rightarrow \Sigma.
\end{equation}
For a fixed $z$, our spectral curve gave $N$ roots for $v$ and $w$, and 
the ramification points are those places where the roots are degenerate, which can be found 
using the spectral curves for two individual Higgs fields.
The genus can be computed using Riemann-Huwitz theorem: 
\begin{equation}
2-2g_{S}=N(2-2g_{\Sigma})-\sum_{p}(e(p)-1)\rightarrow g_{S}={1\over 2}\sum_{p}(e(p)-1)-(N-1)+Ng_{\Sigma},
\end{equation}
here $e(p)$ is the ramification index at point $p$, and there are finite number of ramification points. 

Again, if the spectral curve is only nonzero for one Higgs field , we can find the genus explicitly. 
Let's start with the spectral curve of one Higgs field $\Phi_1\in H^0(\Sigma, End(E)\bigotimes L)$:
\begin{equation}
v^N+\sum_i\phi_i v^{N-i}=0;
\end{equation}
The genus of this Riemann surface is given by the following formula:
\begin{equation}
d_f=\sum_{punctures}p_i+ deg(L){N(N-1)\over 2}+(g-1)(N-1);
\label{fibre}
\end{equation}
here $p_i$ is the contribution from the punctures which is the same as the one contributing the dimension of the moduli space. The difference between the dimensions are (again we assume $deg(L)\geq 2g-2$): 
\begin{equation}
d_b-d_f=(N-1)(deg(L)-(2g-2));
\end{equation}
In $\mathcal{N}=2$ case ($deg(L)=2g-2$), the dimension of the fibre is equal to the dimension of the base, so 
we immediately know the number of  massless photons and the theory is in abelian Coulomb phase if the dimension 
of the moduli space is nonzero.
In $\mathcal{N}=1$ case, the dimension of the fibre is no-longer equal to the dimension of the base, and this leads to many  interesting phases:

\textbf{Non-abelian Coulomb phase}: If there are only regular punctures, then one can define a $C^*$ action on the generalized Hitchin moduli space, and this symmetry can be identified with 
the $U(1)_R$ symmetry of field theory. There is a point on the moduli space which is invariant under this symmetry, and this implies that the theory is conformal.
This phase typically appears at the most singular point (origin) of the moduli space. The property of these new $\mathcal{N}=1$ SCFTs
can be studied in many details using the geometry of M5 branes \cite{xie2014A}.

\textbf{Abelian Coulomb phase}: If the dimension of the fibre $d_b$ is nonzero, then there are massless $U(1)$ photons in the infrared, and this phase is called abelian Coulomb phase. The complex structure 
of the spectral curve is then the exact low energy coupling for the photons \cite{Intriligator:1994sm}. On the Coulomb branch, the singularity points at which the spectral curve is singular is particularly interesting:
at co-dimensional one singularity, there are new massless monopoles; at higher order singularity, we could have Argyres-Douglas points. Given the method of writing the spectral curve, we 
could answer many such questions \cite{xie2014D}. 

\textbf{Confining/Higgs phase}: In some cases, the base of the spectral curve has only finite number of points, and the genus of the spectral curve is zero. This implies  that the theory
could be in confining or Higgs phase. 

\textbf{SUSY breaking}: If we could not find any solution to the constraints, then the supersymmetry might be broken (we have to check the deformation in other direction, which will be considered in next section).

\section{Full moduli space of vacua}

The spectral curve discussed in last section only captures part of moduli space of vacua. 
In fact, 6d $(2,0)$ theory has five scalars which can be used to deform the four dimensional theory. The spectral curve describes 
the deformation associated with four scalars, and there is one more real scalar $\phi_5$ which can be combined with the scalar from gauge fields in Riemann surface
direction to form another complex scalar $\Phi_3$ which is in trivial bundle of Riemann surface, and we can now consider a generalized Hitchin equation with three complex Higgs fields:
\begin{equation}
\Phi_1\in H^{0}(\Sigma, End(E)\bigotimes L_1),~~\Phi_2\in H^{0}(\Sigma, End(E)\bigotimes L_2),~~\Phi_3\in H^{0}(\Sigma, End(E)\bigotimes {\cal O});
\end{equation}
and they commute with each other.  One can now write down the spectral curves for these three scalars following the same method in the last section: simply 
write down the generalized Cayley-Hamilton equation for three pairs $(\Phi_1, \Phi_2)$, $(\Phi_1, \Phi_3)$ and $(\Phi_2, \Phi_3)$. 

\begin{center}
\begin{figure}[htbp]
\small
\centering
\includegraphics[width=10cm]{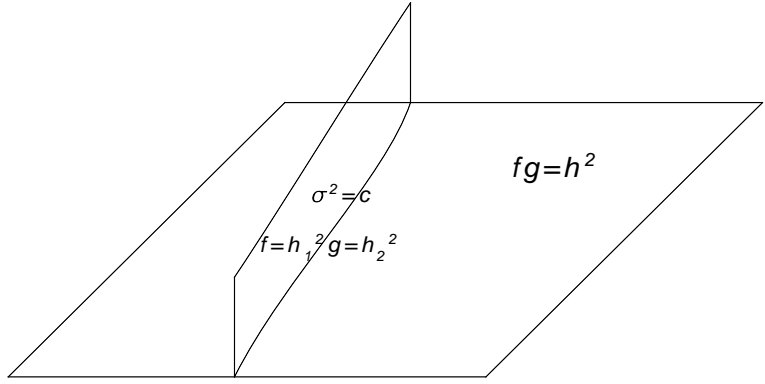}
\caption{Moduli space of vacua of $\mathcal{N}=1$ class ${\cal S}$ theory from 6d $A_1$ theory.}
\label{branch}
\end{figure}
\end{center}

For $A_1$ theory, we have the following equations and constraints for the spectral curves: 
\begin{align}
&v^2=f(z),~~vw=h(z),~~w^2=g(z), \nonumber\\
&v^2=f(z),~~v\sigma=h_1(z),~~\sigma^2=c, \nonumber\\
&w^2=g(z),~~w\sigma=h_2(z),~~\sigma^2=c. 
\label{mixed}
\end{align} 
Here $c$ is just a constant. The commuting relations imply $h^2=f g,~h_1^2=f c,~~h_2^2= g c$, and 
this simply implies that $f, g$ are square of holomorphic sections. The full moduli space is depicted in figure.\ref{branch}.
There might be new branches if there are regular singularities, and the story is quite similar to $\mathcal{N}=2$ case dealt in \cite{Xie:2014pua}.
When there are irregular singularities, the factorization of $f, g$ might not be possible, and we can not turn on $\sigma$ deformations. 

For the general case, let's consider the pair of Higgs fields $(\Phi_1, \sigma)$.  The spectral curve and constraints imply the holomorphic factorization of 
the spectral curve of $\Phi_1$:
\begin{align}
& (v^{n_1}+  h_{1,1} v^{n_1-1}+\ldots+h_{1,n_1})\ldots(v^{n_r}+  h_{r,1} v^{n_1-1}+\ldots+h_{r,n_r})=0, \nonumber \\
& (\sigma-c_1)^{n_1}\ldots(\sigma-c_r)^{n_r}=0.
\end{align}
Here $h_{i,j}$ are the holomorphic sections of various line bundles.  Similar factorization will be applied to the spectral curve of $\Phi_2$. 

\subsection{Examples}
In this subsection, we  will discuss the structure of moduli space of vacua of several interesting examples. Some part of  moduli space of vacua 
has been analyzed in \cite{Xie:2013rsa}, here we describe the full moduli space of vacua.

\subsubsection{Maldacena-Nunez theory}

Maldacena-Nunez theory is defined by the following data: a genus $g$ Riemann surface, and the bundle 
is $K^{1/2}\bigoplus K^{1/2}$. The underlying Riemann surface is chosen to be hyperelliptic,  which is 
described by the following equation:
\begin{equation}
y^2=\prod_{k=1}^{2g+2}(z-p_k),~~~~p_k\neq p_j;
\end{equation} 
The basis for degree one holomorphic differential (holomorphic sections of the canonical bundle) is
\begin{equation}
e_j={z^j dz \over y},~~~j=0,\ldots,g-1;
\end{equation}
The basis for degree $k$ differential is 
\begin{equation}
{z^j dz^k \over y^k},~~~j=0,\ldots, (2 g - 2) k - g. 
\end{equation}

Let's first consider $SU(2)$ theory, and 
the curve is 
\begin{align}
& v^2= \sum_{i=0}^{g-1} v_i e_i, ~~w^2=\sum_{i=0}^{g-1} u_i e_i,~~v w= \sum_{i=0}^{g-1} h_i e_i;
\end{align}
Here $e_i$ is the basis of the holomorphic sections of line bundle $K$ (and the dimension of this space is g). The consistency conditions on the three equations are
\begin{align}
& (\sum_{i=0}^{g-1} h_i e_i)^2= (\sum_{i=0}^{g-1} v_i e_i)(\sum_{i=0}^{g-1} u_i e_i)\rightarrow \sum_{i+j=p} h_i h_j= \sum_{i+j=p} v_i u_j,~~~p=0,1,\ldots, 2g-2;
\label{MN}
\end{align}
Here we use the fact that $e_ie_j=e_{i^{'}}e_{j^{'}}$ if $i+j=i^{'}+j^{'}$. This is the chiral ring relations we are looking for. This example has been considered in \cite{Xie:2013rsa}, in which 
we treat $u_i, v_i$ as special moduli fields. Our new point of view is that the moduli fields in $f, g, h$ are of equal footing, and the moduli space is defined by the above equations. It would 
be interesting to learn more about this moduli space. 

Now let's try to turn on the deformations in $\sigma$, and according to our formula in (\ref{mixed}), this branch is parameterized by the following fields
\begin{equation}
f=h_1^2,~~g=h_2^2,
\end{equation}
where $h_1 \in H^0(\Sigma, K^{1/2})$ and $h_2 \in H^0(\Sigma, K^{1/2})$, and there is an extra moduli field $c$.

\subsubsection{SQCD with quartic superpotential}
Let's consider $\mathcal{N}=1$ $SU(N)$ gauge theory with $N_f\leq 2N-1$ flavors in fundamental representations. We divide the matter into two sets $(n_1,n_2)$ with condition $n_1, n_2<N$.   
Let's use $(q_{i\alpha},  q^{\wedge}_{\alpha i}),~i=1,2,\ldots, n_1$ to denote the first $n_1$ flavor, here $i$ is the flavor index and $\alpha$ is the gauge index,  similarly we use $(Q_{k\alpha}, Q_{\alpha k}^{\wedge}),~k=1,2,\ldots, n_2,$ to
denote the second $n_2$ flavor.  We also add a quartic superpotential for these two sets of flavors:
\begin{equation}
W=c \Tr(\mu_1\mu_2),
\end{equation}
here $\mu_1$ and $\mu_2$ are the moment maps for $SU(N)$ gauge group:
\begin{equation}
(\mu_1)_{\alpha\beta}= q^{\wedge}_{\alpha i}q_{i\beta}-{1\over N}q^{\wedge}_{\gamma i}q_{i\gamma}\delta_{\alpha\beta},
~~~~(\mu_2)_{\beta\alpha}= Q^{\wedge}_{\beta k}Q_{k\alpha}-{1\over N}Q^{\wedge}_{\gamma k}Q_{k\gamma}\delta_{\beta\alpha}.
\end{equation}
Let's split the meson matrix as 
\begin{equation}
M=\left(\begin{array}{cc}
M_1& L_1\\
L_2& M_2
\end{array}\right ),
\end{equation}
with 
\begin{align}
&(M_1)_{ij}=q_{i\alpha} q_{\alpha j}^{\wedge},~i,j=1,2,\ldots, n_1,~~~~(M_2)_{kl}=Q_{k\alpha} {Q}_{\alpha l}^{\wedge},~k,l=1,2,\ldots, n_2 \nonumber\\
& (L_1)_{ik}=q_{i\alpha} Q_{\alpha k}^{\wedge},~~~~~~~~~~~~~~~~~~~~~~~~~~~(L_2)_{ki}=Q_{k\alpha}q_{\alpha i}^{\wedge}.
\end{align}
Then the above quartic superpotential reads
\begin{equation}
W=c tr(\mu_1\mu_2)=c\Tr (L_1 L_2)-{c\over N}\Tr(M_1)\Tr(M_2);
\end{equation}
This is the tree level quartic superpotential added for SQCD. To find out the moduli space, we need to consider the quantum generated 
superpotential. Here we will use M5 brane construction and the spectral curve method to determine the full moduli space, and check with
the field theory method.

\begin{center}
\begin{figure}[htbp]
\small
\centering
\includegraphics[width=12cm]{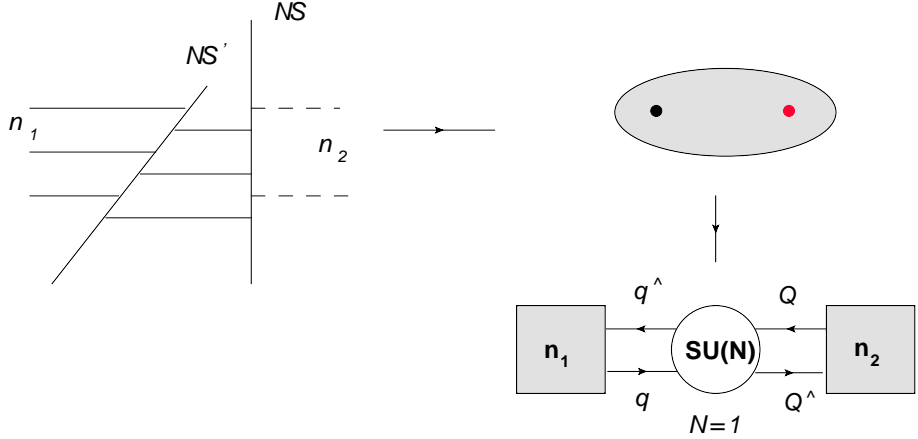}
\caption{Left:~Type IIA brane configuration for SQCD deformed by a quartic superpotential. Right: The lift to M5 brane description: there are two irregular singularities describing  
left and right NS5 branes. }
\label{quart}
\end{figure}
\end{center}

The M5 brane configuration for above SQCD deformed by a quartic superpotential needs irregular punctures. The boundary condition for one irregular puncture representing $n$ flavors are
\begin{equation}
\Phi\sim {\zeta\over z^{1+{1\over N-n}}}\diag(0,\ldots,0,1,\omega,\ldots, \omega^{N-n-1})+\ldots,
\end{equation}
here $\omega^{N-n}=1$. When $n=N-1$, the irregular puncture has the following form
\begin{equation}
\Phi\sim {\zeta\over z^2}\diag(1,1,\ldots, -(N-1))+\ldots,
\end{equation}
The M5 brane configuration for SQCD with $(n_1,n_2)$ flavors with quartic superpotential are described by a sphere with two irregular punctures of above type: $\Phi_1$ is singular at one point, and $\Phi_2$
is singular at another point. We put two punctures at $z=0$ and $z=\infty$ separately. 
 The bundle structure is ${\cal O}(-1)\bigoplus{\cal O}(-1)$. With the above boundary condition and bundle structure, we can solve 
the spectral curve using our general formula. In the following, we will discuss two simple examples. The Type IIA brane configuration and its lift to M5 brane configuration are shown in figure. \ref{quart}.

\textbf{A:~$SU(2)$ with $(1,1)$ flavors revisited}:
The coefficients of the full spectral curves are
\begin{align}
&v^2={\zeta_1^2\over z^4}+{u_1\over z^2},~~v w= {u_3\over z^2}+{u_4\over z}+u_5,~~w^2=\zeta_2^2 z^2 +u_2, \nonumber\\
&\sigma^2=c^2, ~~~\sigma v={u_6\over z^2}+{u_7\over z},~~\sigma w= u_8 z+u_9.
\end{align}
Substituting this into equations (\ref{mixed}), we find two branches
\begin{align}
&\text{Solution 1}: u_1=u_2=u_3=u_5=u_7=u_9=0, ~u_4=\eta \zeta_1^2,~u_6=c \zeta_1,~u_8=c \zeta_2,~~~\eta^2=1,   \nonumber\\
&\text{Solution 2}:c=u_6=u_7=u_8=u_9=0,~~ u_1= {u_5^2\over \zeta_2^2},~~u_2= {u_3^2\over \zeta_1^2},~~u_3 u_5 =\zeta_1^2\zeta_2^2.
\end{align}
The first branch has two components which are called branch 1 and branch $1^{'}$, and there are a total of three branches. 
$\zeta_1$ and $\zeta_2$ can be taken to be equal by using scale transformation on $v, w, z$ coordinates, so there is only one parameter
which can be identified with the dynamical generated scale.

Field theory method has been used in \cite{Xie:2013rsa} to determine the full moduli space of this theory. The tree level superpotential plus
the deformed chiral ring relation constraint is simply
\begin{equation}
W=X(M_{11}M_{22}-M_{12}M_{21}-B\tilde{B}-\Lambda^4)+c(M_{12}M_{21}-{1\over 2} M_{11}M_{22});
\end{equation} 
The flavor symmetry is reduced to $U(1)\times U(1)$, and  $M_{11}, M_{22}$  are flavor invariant meson field. There are three branches
\begin{align}
&\text{Branch A}:~~ M_{12}M_{21}=-\Lambda^4,~~ M_{11}=M_{22}=B=\tilde{B}=0,\nonumber\\
&\text{Branch B}:~~B\tilde{B}=-\Lambda^4,~~ M_{11}=M_{22}=M_{12}=M_{21}=0, \nonumber\\
&\text{Branch C}:~~ M_{11}M_{22}=\Lambda^4,~~ M_{12}=M_{21}=B=\tilde{B}=0.
\end{align}
The branch C can be identified with the branch 2 in our spectral curve: $u_1$ and $u_2$ are flavor invariant moduli fields which are 
just $M_{11}$ and $M_{22}$.  The branch $A$ and $B$ can be identified with the branch 1 and $1^{'}$ in spectral curve picture. There is only one flavor invariant moduli fields
in these two branches, which match with the result from spectral curve.

\textbf{B:~$SU(3)$ with $(1,1)$ flavors}: The coefficients before imposing the chiral-ring relation are
\begin{align}
&f_1={\zeta_1^2\over z^3}+{u_1\over z^2},~~g_1={u_2\over z^3},\nonumber \\
&f_3={u_3\over z},~~~g_3={u_4\over z^3}+{u_5\over z^2},~~~~~g_4={u_6\over z}+u_7,  \nonumber\\
&f_2=\zeta_1^2 z+u_8,~~~g_2=u_9.
\end{align}
Substituting the above coefficient into equations (\ref{conA2}), we find two sets of solutions:
\begin{align}
&\text{Solutions 1}: u_1=u_2=u_4=u_5=u_6=u_7=u_8=u_9=0,~~u_3={2\eta \zeta_1^2\over3},~~\eta^2=1  \nonumber\\
&\text{Solutions 2}:   u_3=\eta \frac{2\zeta_1^2}{3\sqrt{3}},~u_7=-\eta \frac{4 \zeta_1^6}{27 \sqrt{3} u_4},~u_1= -\frac{27 u_7^2}{4 \zeta_1^4}, u_8=-\frac{27 u_4^2}{4 \zeta_1^4}, \nonumber\\
&u_6= \frac{2 \eta u_4}{\sqrt{3}}, u_9= \frac{27 u_4^3}{4 \zeta_1^6}, u_5 = -\frac{8 \zeta_1^6}{81 u_4}, u_2 =-\frac{\eta u_7^2}{\sqrt{3} u_4},~~~~\eta^2=1.
\end{align}
In the first set of solutions, there are two vacua. Since the the curves are factorized holomorphically, we can turn on one dimensional deformation in $\sigma$ direction. 
There are two branches in the second set of solutions, and there are two moduli fields satisfying a chiral ring relation:
\begin{equation}
u_4 u_7 \sim  \eta \zeta^6,~~~\eta^2=1.
\end{equation}

The field theory analysis is easy to find. The full superpotential involving the dynamical generated superpotential is 
\begin{equation}
W=c(L_1 L_2-\frac{1}{3} M_1 M_2)+{\Lambda^{7}\over M_1 M_2 -L_1 L_2}
\end{equation}
The critical points of above potential is easy to find and there are four branches:
\begin{align}
&\text{Branch 1}:~M_1 M_2={i\sqrt{3}\Lambda^{7/2}\over \sqrt{c}},~~~~~L_1=L_2=0,   \nonumber\\
&\text{Branch 2}:~M_1 M_2=-{i\sqrt{3}\Lambda^{7/2}\over \sqrt{c}},~~~~~L_1=L_2=0,   \nonumber\\
&\text{Branch 3}:~L_1 L_2={i\Lambda^{7/2}\over \sqrt{c}},~~~~~M_1=M_2=0, \nonumber\\
&\text{Branch 4}:~L_1 L_2=-{i\Lambda^{7/2}\over \sqrt{c}},~~~~~M_1=M_2=0. 
\end{align}
Here $M_1$ and $M_2$ are flavor invariant meson, so they can be the moduli fields, so branch 1 and 2 are identified with the solutions set 2 (we do not attempt to match the constants).
 For the other two branches, there is only one flavor invariant moduli fields, which are identified with the solutions set 1. Some of the branches are missed in \cite{Xie:2013rsa}, here
we find the full set of vacua using the spectral curve. 

\section{Theories with other superpotentials}
The theory considered in last section has  quartic superpotential. In this section, we will consider other types of superpotential. 

\subsection{Quardratic superpotential}
Let's consider a  pure $\mathcal{N}=2$ $SU(N)$ theory deformed by the following superpotential
\begin{equation}
W=\mu \Tr (\Phi^2),
\end{equation}
here $\Phi$ is the adjoint chiral superfield in $\mathcal{N}=2$ vector multiplet.  The infrared structure is 
the same as the pure $\mathcal{N}=1$ theory: there are only $N$ vacua. 

\begin{center}
\begin{figure}[htbp]
\small
\centering
\includegraphics[width=15cm]{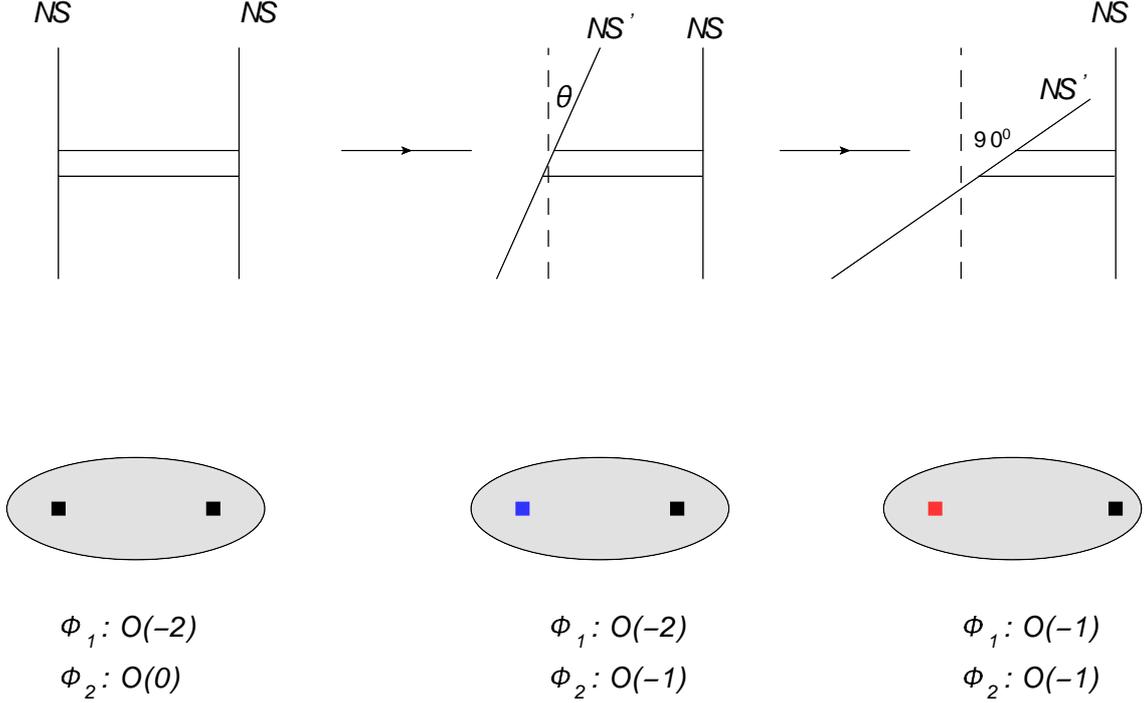}
\caption{Left: Type IIA and M5 brane configuration for $\mathcal{N}=2$ theory, here the two irregular punctures are of the same type. Middle: Type IIA and M5 brane configuration for  $\mathcal{N}=2$ theory deformed by finite
adjoint mass, here $\Phi_1$ and $\Phi_2$ are both singular at the rotated puncture labeled by a blue square; Right: Type IIA and M5 brane configuration for  pure $\mathcal{N}=1$ theory, here $\Phi_1$ and $\Phi_2$ 
are only singular at one puncture respectively. }
\label{adj}
\end{figure}
\end{center}

The corresponding type IIA
brane configuration is found by rotating one of NS5 brane by an angle $\theta$, the adjoint mass is given by
\begin{equation}
\mu\sim \tan \theta. 
\end{equation}
When $\theta=90^0$, we get pure $\mathcal{N}=1$ theory. The type IIA configurations and M5 brane configuration are shown in figure.~\ref{adj}. 
Since $\Phi_1$ ($\Phi_2$) is describing the deformation in $v$ ($w$) direction, when the rotated angles are not 90 degrees, we claim that 
both $\Phi_1$ and $\Phi_2$ are singular at the singularity representing the rotated brane. At the other singularity representing the unrotated
brane, only $\Phi_1$ is singular.  The bundle structure is fixed by the $\mathcal{N}=1$ condition: 
\begin{equation}
d_1+n_1\geq0,~~d_2+n_2\geq0,~~d_1+d_2+n_1+n_2=-2+n,
\end{equation}
here $d_i$ is degree of the line bundle $L_i^{'}$, and $n_i$ is the number of punctures of $\Phi_i$, and $n$ is total number of punctures.
 Here we have $n_1=2,n_2=1,n=2$, and we find the unique solution
\begin{equation}
d_1=-2,~~d_2=-1.
\end{equation}

The singular behavior at the punctures for both Higgs fields are
\begin{align}
&\Phi\sim{\zeta_1\over z^{1+1/2}}\diag(1,-1), \nonumber \\
\end{align}
here $\Phi_1$ is singular at infinity and zero, and the bundle structure is ${\cal O}(-2)$. $\Phi_2$ is only singular at  infinity, and is a section of bundle ${\cal O}(-1)$. Based on above data, we can easily find the 
spectral curve
\begin{align}
&v^2={\zeta_1^2  \over z^3}+{u \over  z^2} +{\zeta_2^2\over z},~~w^2=\zeta_3^2 z+u_1,~~vw =h(z).
\end{align}
and the constraints equation is
\begin{equation}
h^2=({\zeta_1^2  \over z^3}+{u \over  z^2} +{\zeta_2^2\over z})(\zeta_3^2 z+u_1 ).
\end{equation}
To make $h$ holomorphic, we have to set 
\begin{equation}
u=2\eta\zeta_1\zeta_2,~~~u_1=0.
\end{equation}
In summary, the solution we found is 
\begin{align}
&v^2={1  \over z^3}(\eta \zeta_2 z+\zeta_1)^2,~w^2=\zeta_3^2 z,~~ vw = \zeta_3 (\eta\zeta_2 z+\zeta_1)/z,
\end{align}
with $\eta^2=1$, so there are two vacua. Notice that the equation for $v$ is the SW curve of $\mathcal{N}=2$ theory, and the value of $u$ determined by spectral curve is 
exactly the position where the SW curve is degenerate. It is pointed in \cite{Seiberg:1994rs} that only these two points survive with the adjoint mass deformations. Our curve 
reproduces this result. 

 Since the above curve can not be factorized in a holomorphic way, $\sigma$ deformation can not be turned on, those are the only two vacua. This is 
apparently in agreement with the field theory result.

The generalization to SU(N) gauge theory is straightforward.
We only work  out mass deformed $\mathcal{N}=2$ $SU(3)$ theory and leave the general case for the interested reader. 
The M5 brane construction is similar as $SU(2)$ theory: there are two irregular singularities at $z=0,\infty$ with the following behavior for the Higgs field:
\begin{equation}
\Phi\sim{\zeta\over z^{1+1/3}}\diag(1, \exp(i2\pi/3),\exp(i4\pi/3))+\ldots 
\end{equation} 
$\Phi_1$ is singular at both points while $\Phi_2$ is only singular at $z=\infty$. Again, the bundle structure is ${\cal O}(-2)\bigoplus {\cal O}(-1)$. 
Based on above data, the coefficients in the spectral curve reads:
\begin{align}
&f_1={u_1\over z^2},~~g_1={{\zeta_1^3}\over z^4}+{u_2\over z^3}+{\zeta_2^3\over z^2}, \nonumber\\
&f_3={u_3\over z},~~g_3= {u_4\over z^2}+{\zeta_2^2\zeta_3\over z},~~ g_4={u_5\over z}+{\zeta_2\zeta_3^2},      \nonumber\\
&f_2=u_6,~~~g_2=\zeta_3^2 z+ u_7.
\end{align}
Substitute the above form into our constraints equations (\ref{conA2}), we find
\begin{equation}
u_2=u_4=u_5=u_6=u_7=0,~~~~~u_1=3\eta \zeta_1\zeta_2,~~~u_3=\eta \zeta_1\zeta_3,
\end{equation}
with $\eta^3=-1$; and the final curve is 
\begin{align}
&v^3+{3\eta \zeta_1\zeta_2 \over z^2}v +{\zeta_1^3\over z^4}+{\zeta_2^3\over z^2}=0, \nonumber\\
&v^2 w+{\eta \zeta_1\zeta_2 \over z} w+{\eta \zeta_1\zeta_3 \over z} v+{\zeta_2^2\zeta_3\over z}=0, \nonumber\\
&vw^2 +{\eta \zeta_1\zeta_3  \over z} w+{\zeta_2\zeta_3^2}=0, \nonumber\\
&w^3+\zeta_3^3 z=0.
\end{align}
So we find three vacua (the deformation in $\sigma$ direction is not possible),
 and the result is in agreement with the field theory result found in \cite{Argyres:1995jj}. A further check is  to note that the value of $u_1$ 
and $u_2$ are exactly the points on $\mathcal{N}=2$ moduli space where two mutually local monopoles become massless, which are the unlifted $\mathcal{N}=1$ vacua after turning on the adjoint mass deformation. Let's give more detail on this point. The 
curve of $v$ can be regarded as the $\mathcal{N}=2$ curve: after changing coordinates $x=v z$, we have 
\begin{align}
&v^3+{u_1\over z^2} v+{{\zeta_1^3}\over z^4}+{u_2\over z^3}+{\zeta_2^3\over z^2}=0\rightarrow x^3+u_1 x+u_2+\zeta^3/z+\zeta^3 z=0 \nonumber\\
&\rightarrow y^2=[{1\over 2}(x^3+u_1 x+u_2)]^2-\zeta^6.
\end{align}
We have used the scale invariance to put $\zeta_1=\zeta_2=\zeta$ and $y$ is a linear function in $z$. The last equation is the standard curve found in \cite{Argyres:1994xh}, so the moduli $u_1$ and 
$u_2$ in our formula are the same as the one used in the old $\mathcal{N}=2$ literature. The value of the moduli fields at the unlifted vacua upon $\mathcal{N}=1$ deformations 
are exactly the same as found in  \cite{Argyres:1994xh}.

\subsection{Landau-Ginzburg superpotential}
Let's now consider the $\mathcal{N}=2$ pure SU(N) gauge theory deformed by following Landau-Ginzburg (LG) type superpotential:
\begin{equation}
W=\sum_{i=2}^{k}\mu_i\Tr(\Phi^i),~~~~~k\leq N.
\end{equation}
The vacua structure of this deformed theory is very rich! See for example \cite{deBoer:1997ap, Dijkgraaf:2002fc, Cachazo:2002ry, Cachazo:2002zk}.
The type IIA brane construction for this superpotential is suggested in \cite{Elitzur:1997hc}: they argue that 
one need to use coincident multiple NS5 branes. The M5 brane configuration and $\mathcal{N}=1$ curve is also  discussed in \cite{deBoer:1997ap}. 

\begin{center}
\begin{figure}[htbp]
\small
\centering
\includegraphics[width=8cm]{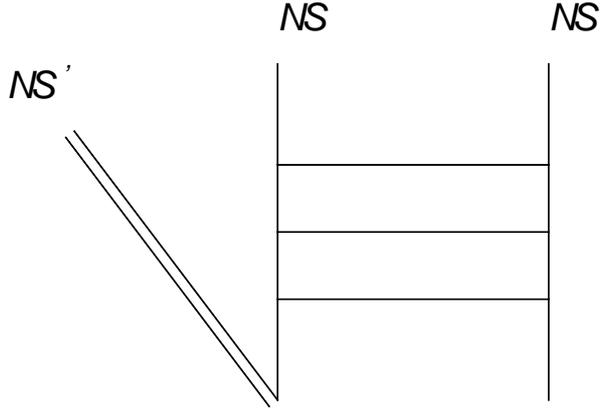}
\caption{Type IIA brane configuration for $\mathcal{N}=2$ theory deformed by LG superpotential.}
\label{LG}
\end{figure}
\end{center}

However, the true type IIA brane configuration seems slightly different: the branes on the left hand side might be interpreted as one NS5 brane in original 
direction, and k $NS5^{'}$ aligned in orthogonal direction, see figure. \ref{LG}. Based on this conjecture, the M5 brane configuration for 
 $N=2$ $SU(N)$ gauge theory deformed by LG superpotential can be engineered 
by a sphere with following boundary conditions: 
\begin{align}
& \Phi_1\sim   {\zeta_1\over z^{'1+{1\over N}}}(1,\omega,\ldots, \omega^{N-1}),~~z\rightarrow \infty,~~~~   \Phi_1\sim   {\zeta_2\over z^{1+{1\over N}}}(1,\omega,\ldots, \omega^{N-1})                     ~~z\sim0, \nonumber\\
& \Phi_2 \sim {\zeta_3\over z^{'1+{1\over N}}}(1,\omega,\ldots, \omega^{N-1}),  ~~z\sim \infty,
\end{align}
here $z^{'}=1/z$, $\omega^N=1$. The bundle structure is $O(-2)\bigoplus O(-1)$.  Let's now find the curves for deformed $\mathcal{N}=2$ $SU(3)$ pure YM theory, and consider the following potential
\begin{equation}
W=u \Tr(\Phi^3)+v \Tr(\Phi^2),
\end{equation}
here $\Phi$ is the adjoint scalar in $\mathcal{N}=2$ vector multiplet. The case with $u=0$ has been studied in last subsection, and the case $u=0, v=\infty$ has been
studied in \cite{Xie:2013rsa} (one can also recover this result using the method presented in this paper).

Let's first consider the special situation $v=0$, which means that we only turn on leading order term of $\Phi_2$.  Using the boundary data and the bundle structure, 
the coefficients of the spectral curve have the following form:
\begin{align}
&f_1={u_1\over z^2},~~g_1={{\zeta_1^3}\over z^4}+{u_2\over z^3}+{\zeta_1^3\over z^2}, \nonumber\\
&f_3={u_3\over z}-\zeta_1\zeta_3,~~g_3={u_4\over z^2}+{u_5\over z},~~g_4= {u_6\over z}+u_7,~~      \nonumber\\
&f_2=u_8,~~~g_2=\zeta_3^3z^2+u_9 z+u_{10}.
\end{align}
 Substituting the above form into our constraints equations, we have  two sets of solutions:
 \begin{align}
 & \text{Solutions 1}:~u_2=u_3=u_5=u_6=u_8=u_9=u_{10}=0,~~~~\nonumber\\
  &  ~~~~~~~u_1=3\eta\zeta_1^2,~~u_4=\eta^2 \zeta_1^2\zeta_3,~~u_7=-\eta \zeta_1\zeta_3^2,~\eta^3=-1, \nonumber\\
 &\text{Solutions 2}:~u_1=u_4=u_5=u_6=u_7=u_8=u_{10}=0,~~\nonumber\\
 &~~~~~~~~u_2=2\eta  \zeta_1^3,~~u_3=-\eta \zeta_1\zeta_3,~~u_9=\eta \zeta_3^3,~\eta^2=1.
 \end{align}
For the first branch, the value of $u_1$ and $u_2$ are the position where there are two mutually local massless monopoles. 
For the second branch, the value of $u_1$ and $u_2$ are the Argyres-Douglas point found in \cite{Argyres:1995jj}. 
In total, we have found five vacua and these are precisely the vacua found in \cite{Argyres:1995jj} in the case of turning on $\Tr(\Phi^3)$ deformation on pure $\mathcal{N}=2$ $SU(3)$ gauge theory. 
The deformation in $\sigma$ direction is not possible, so these are the isolated vacua.

For  general $u, v$ deformation, we need to turn on lower order term of $\Phi_2$. The coefficients in the spectral curve have the following form 
\begin{align}
&f_1={u_1\over z^2},~~g_1={{\zeta_1^3}\over z^4}+{u_2\over z^3}+{\zeta_1^3\over z^2}, \nonumber\\
&f_3={u_3\over z}-\zeta_1\zeta_3,~~g_3={u_5\over z^2}+{u_6\over z},~~g_4= {u_7\over z}+u_8,      \nonumber\\
&f_2=d z+u_9,~~~g_2=\zeta_3^3z^2+u_{10} z+u_{11}.
\end{align}
The only difference with $v=0$ case is that $f_2$ has a order $z$ term. The parameters in the spectral curve and the parameters of the physical theory are identified as:
\begin{equation}
\Lambda=\zeta_1,~~~u=\zeta_3,~~~v={d\over 3 \zeta_3}.
\end{equation}

Substitute above coefficients into equations (\ref{conA2}), we find the following two sets of solutions:
\begin{align}
&\text{Solutions 1}:~~u_2=u_7=u_9=u_{11}=0,~u_1=-3 \eta^2 \zeta_1^2,~ u_3={\zeta_1 d \over 3 \zeta_3 \eta},~u_5=\eta \zeta_1^2\zeta_3, u_6=-{\zeta_1^2 d\over 3 \zeta_3} \nonumber\\
&~~~~~~~~~~~~~~~~u_8={\zeta_1^2 d^2+9 \zeta_1^2\zeta_3^2\eta^2\over 9\zeta_1\zeta_3^2},~~u_{10}=-{d^3\over 27 \zeta_3^3},~~~~\eta^3=1; \nonumber\\
&\text{Solutions 2}:~~u_7=u_9=u_{11}=0,~u_1=-{\zeta_1^2 d^2 \over 3 \zeta_3^4},~u_2={2 \zeta_1^3 d^3\over27 \zeta_3^6} - 2\eta \zeta_1^3,~~ u_3={\zeta_1\zeta_3\over \eta},
u_5={\eta \zeta_1^2 d\over 3\zeta_3}, \nonumber\\
&~~~~~~~~~~~~~~~~~u_6=-{\zeta_1^2 d\over 3 \zeta_3},~u_8 =-{2 \zeta_1 d^2\over9 \zeta_3^2},~u_{10}=-\eta \zeta_3^3,~~~~\eta^2=1.\nonumber\\
\end{align}
For the first set of solutions, $u_1$ and $u_2$ are exactly the value where there are two mutually local massless monopoles, and there are three vacua. For general parameters, the
second set of solutions have only two vacua. So there are again a total of five vacua ($\sigma$ deformation is not possible). The phase structure can be found by calculating the 
genus of the spectral curve: for general $u, v$, the spectral curve corresponds to first three vacua has genus zero, and therefore there is no massless photon, and the vacua is gapped.
For the second set of vacua, the genus of the curve is one, and there is one massless photon.

There are several special values of $u, v$ so that one of the vacua in second set will merge with the first set of vacua
\begin{align}
&\eta=1:~~d=3\gamma \zeta_3^2 (v=\gamma u),~~\gamma^3=1\rightarrow u_1=  -3 \zeta_1^2 \gamma^2,~u_2=0,   \nonumber\\
&\eta=-1:~~d=-3\gamma \zeta_3^2 (v=-\gamma u),~\gamma^3=1~\rightarrow u_1=  -3 \zeta_1^2 \gamma^2,~u_2=0.
\end{align}
So there are six special ratio of $(u,v)$ for which one of vacua in second set is merged with one of the vacua in first set. 

In summary, we have found the following intricate vacua structure of $SU(3)$ gauge theory:
\begin{itemize}
\item $u=0, v=0$: The theory has $\mathcal{N}=2$ SUSY, and there is a $\mathcal{N}=2$ Coulomb branch.
\item $u=0, v\neq 0$: there are three vacua: they are points of $\mathcal{N}=2$ Coulomb branch   with  two mutually local massless monopoles . 
The genus of the spectral curve is zero, so there is no massless photon, and 
the theory is in confining phase. 
\item $u=0, v=\infty$: there are three vacua, and they are in confining phase.
\item $u\neq 0,v=0$: there are five vacua: three of them are points of $\mathcal{N}=2$ Coulomb branch   with  two mutually local massless monopoles, and they are gapped; the other two are $\mathcal{N}=2$ AD points deformed by a superpotential . 
\item $v/u=\pm \gamma, \gamma^3=1$:  three are four vacua: two vacua are points with two mutually local massless monopoles; the third vacua has one massless photon, and  the fourth one has one massless photon and one massless hypermultiplet.
\item $v, u$ generic: there are five vacua: three of them are points of $\mathcal{N}=2$ Coulomb branch   with  two mutually local massless monopoles, and they are gapped; the other two are half-Higgsed and there is one massless photon left
at each vacua.
\end{itemize}
These are precisely the vacua found in \cite{Argyres:1995jj}, and we found a curve for all the cases.
 It is remarkable that our spectral curve recovers these highly non-trivial vacua structure without using any field theory input!

 \section{Conclusion}
In this paper, we propose a $\mathcal{N}=1$ curve for rank N  four dimensional class ${\cal S}$ theory with at least $\mathcal{N}=1$ supersymmetry. The curve consists of three parts:
a: a set of N+1 equations; b: the constraints relating the coefficients which are holomorphic sections of various line bundles; c: a canonically defined 
differential.  Therefore we establish a Seiberg-Witten type solution for $\mathcal{N}=1$ theory. 
The big difference with $\mathcal{N}=2$ curve is the constraints on coefficients which lead to many new phenomenon of 
$\mathcal{N}=1$ theory such as more phases, chiral ring relations for moduli fields, etc. 

We have applied this method to some known theories, and they recover the intricate vacua structure in an impressive way. 
The main purpose is to check the correctness of our proposal, so we mainly focus on examples where other methods 
 are available. Given the compelling evidence for the correctness of our proposal, we can apply it to all kinds of new $\mathcal{N}=1$ theories engineered using M5 branes including 
SCFT, asymptotical free theory, Argyres-Douglas type theories \cite{xie2014A,xie2014B}, etc.  The spectral curve is a important tool to understand the 
 properties of those new theories as other methods are missing. 
 
 There are some remaining questions about the constraints of our spectral curve. For example, the constraints are easy to derive and have nice pattern, 
 but kind of complicated for higher rank theory. Is there any way to simplify them?
  
 We only discussed how to write down the curve, and it is definitely interesting to analyze these curves in details such as the singularity structure, phase diagram, etc. 
 We believe that many exact results  about $\mathcal{N}=1$ theories can now be tackled by using the curve presented in this paper. 
 
Our curves are derived  using M5 brane method, and it would be interesting to see if they can be derived using other string duality such as the mirror symmetry of
type II string theory. The construction in \cite{Dijkgraaf:2002fc} seems closely  related to ours: they also use the dimensional reduction of 
holomorphic Chern-Simons action (our generalized Hitchin equation is also derived from dimensional reduction of 
the same action), although the ways of introducing superpotential are different in those two constructions. 
It would be nice to understand better the relations between these two constructions.

 \acknowledgments
 This work was supported by Center of Mathematical Sciences and Applications at Harvard University,  
 and in part by the Fundamental Laws Initiative of the Center for the Fundamental Laws of Nature, Harvard University.

 \appendix
 
 \section{Spectral curve and constraints for $A_3$ theory}
The spectral curve of $A_3$ theory are derived from order 4 matrix equations of two commuting $4\times 4$ matrices. These matrix equations can be found using the basic Cayley-Hamilton equation:
\begin{equation}
p(\lambda)=\text{det}(\lambda-X)=0\rightarrow X^4-\frac{1}{2}X^2 \text{Tr}\left[X^2\right]-\frac{1}{3}X \text{Tr}\left[X^3\right]+[-\frac{\text{Tr}\left[X^4\right]}{4}+\frac{1}{8} \text{Tr}\left[X^2\right]^2]=0. 
\end{equation}
and we have a total of five degree 4 equations:
\begin{align}
& \Phi_1^4-\frac{1}{2} \Phi_1^2 \text{Tr}\left[\Phi_1^2\right]-\frac{1}{3} \Phi_1 \text{Tr}\left[\Phi_1^3\right]+[-\frac{\text{Tr}\left[\Phi_1^4\right]}{4}+\frac{1}{8} \text{Tr}\left[\Phi_1^2\right]^2]=0; \nonumber\\
& \Phi_1^3 \Phi_2-\frac{1}{4} \Phi_1 \Phi_2 \text{Tr}\left[\Phi_1^2\right]-\frac{1}{4} \Phi_1^2 \text{Tr}[\Phi_1 \Phi_2]-\frac{1}{12} \Phi_2 \text{Tr}\left[\Phi_1^3\right] 
 -\frac{1}{4} \Phi_1 \text{Tr}\left[\Phi_1^2 \Phi_2\right]+ \nonumber\\
 & \left[-\frac{1}{4} \text{Tr}\left[\Phi_1^3 \Phi_2\right]+\frac{1}{8} \text{Tr}\left[\Phi_1^2\right] \text{Tr}[\Phi_1 \Phi_2]\right]=0; \nonumber\\
&\Phi_1^2 \Phi_2^2-\frac{1}{12} \Phi_2^2 \text{Tr}\left[\Phi_1^2\right]-\frac{1}{3} \Phi_1 \Phi_2 \text{Tr}[\Phi_1 \Phi_2]-\frac{1}{6} \Phi_2 \text{Tr}\left[\Phi_1^2 \Phi_2\right]-\frac{1}{12} \Phi_1^2 \text{Tr}\left[\Phi_2^2\right]-\frac{1}{6} \Phi_1 \text{Tr}\left[\Phi_1 \Phi_2^2\right]+ \nonumber\\
&\left[-\frac{1}{4} \text{Tr}\left[\Phi_1^2 \Phi_2^2\right]+\frac{1}{12} \text{Tr}[\Phi_1 \Phi_2]^2+\frac{1}{24} \text{Tr}\left[\Phi_1^2\right] \text{Tr}\left[\Phi_2^2\right]\right]=0; \nonumber\\
&\Phi_2^3 \Phi_1-\frac{1}{4} \Phi_2 \Phi_1 \text{Tr}\left[\Phi_2^2\right]-\frac{1}{4} \Phi_2^2 \text{Tr}[\Phi_2 \Phi_1]-\frac{1}{12} \Phi_1 \text{Tr}\left[\Phi_2^3\right]-\frac{1}{4} \Phi_2 \text{Tr}\left[\Phi_2^2 \Phi_1\right]+ \nonumber\\
&\left[-\frac{1}{4} \text{Tr}\left[\Phi_2^3 \Phi_1\right]+\frac{1}{8} \text{Tr}\left[\Phi_2^2\right] \text{Tr}[\Phi_2 \Phi_1]\right]=0; \nonumber\\
 & \Phi_2^4-\frac{1}{2} \Phi_2^2 \text{Tr}\left[\Phi_2^2\right]-\frac{1}{3} \Phi_2 \text{Tr}\left[\Phi_2^3\right]+\left[\frac{1}{8} \text{Tr}\left[\Phi_2^2\right]^2-\frac{\text{Tr}\left[\Phi_2^4\right]}{4}\right]=0.
 \end{align}
The above equation can be derived by the following steps: first list all  monomials with $a$ $\Phi_1$ and $(4-a)$ $\Phi_2$ factors with undetermined coefficients, then use two diagonal matrices 
to fix the coefficients. The spectral curve is written using the above generalized trace identities:
\begin{align}
&v^4+f_{2,0} v^2+f_{3,0} v+f_{4,0}=0, \nonumber\\
&v^3 w +v^2 f_{1,1} + v w {f_{2,0}\over 2}+v f_{2,1}+ w {f_{3,0}\over 4}+f_{3,1}=0, \nonumber\\
&v^2w^2+v^2 {f_{0,2}\over 6}+w^2 {f_{2,0}\over 6}+v w{4 f_{1,1}\over3}+ w {2 f_{2,1}\over 3}+v {2f_{1,2}\over 3}+f_{2,2}=0,  \nonumber\\
&vw^3 +w^2 f_{1,1} + v w {f_{0,2}\over2}+w f_{1,2}+ v {f_{0,3}\over 4}+f_{1,3}=0,  \nonumber\\
&w^4+f_{0,2} v^2+f_{0,3} v+f_{0,4}=0.  
\label{curA3}
\end{align}
and the coefficients are expressed in terms of traces of $\Phi_1$ and $\Phi_2$:
\begin{align}
& f_{2,0}=-\frac{1}{2}\text{Tr}[\Phi_1^2],~f_{3,0}=-\frac{1}{3}\text{Tr}[\Phi_1^3], ~f_{4,0}=-\frac{\text{Tr}[\Phi_1^4]}{4}+\frac{1}{8} \text{Tr}\left[\Phi_1^2\right]^2,   \nonumber\\
& f_{1,1}=-\frac{1}{4}\text{Tr}[\Phi_1\Phi_2],~ f_{2,1}=-{1\over 4}\text{tr}[\Phi_1^2\Phi_2], ~f_{3,1}=-\frac{1}{4} \text{Tr}\left[\Phi_1^3 \Phi_2\right]+\frac{1}{8} \text{Tr}\left[\Phi_1^2\right] \text{Tr}[\Phi_1 \Phi_2],   \nonumber\\      
& f_{1,2}=-{1\over 4}\text{tr}[\Phi_1\Phi_2^2],~ f_{2,2}=-\frac{1}{4} \text{Tr}\left[\Phi_1^2 \Phi_2^2\right]+\frac{1}{12} \text{Tr}[\Phi_1 \Phi_2]^2+\frac{1}{24} \text{Tr}\left[\Phi_1^2\right] \text{Tr}\left[\Phi_2^2\right],  \nonumber\\
& f_{1,3}=-\frac{1}{4} \text{Tr}\left[\Phi_2^3 \Phi_1\right]+\frac{1}{8} \text{Tr}\left[\Phi_2^2\right] \text{Tr}[\Phi_2 \Phi_1],     \nonumber \\
& f_{0,2}=-\frac{1}{2}\text{Tr}[\Phi_2^2],~ f_{0,3}=-\frac{1}{3}\text{Tr}[\Phi_2^3],~ f_{0,4}=-\frac{\text{Tr}[\Phi_2^4]}{4}+\frac{1}{8} \text{Tr}\left[\Phi_2^2\right]^2.
\end{align}

The constraints for the coefficients can be easily derived from the matrix equations, and we have:
\begin{align}
&\text{Type}[4,2]:~-\frac{f_{0,2} f_{2,0}^2}{3}+2 f_{0,2} f_{4,0}+\frac{4 f_{1,1}^2 f_{2,0}}{3}-8 f_{1,1} f_{3,1}+3 f_{1,2} f_{3,0}+2 f_{2,0} f_{2,2}-4 f_{2,1}^2=0; \nonumber\\
&\text{Type}[3,3]:~~-\frac{4 f_{0,2} f_{1,1} f_{2,0}}{9}+\frac{4 f_{0,2} f_{3,1}}{3}+\frac{3 f_{0,3} f_{3,0}}{4}+\frac{16 f_{1,1}^3}{9}-\frac{16 f_{1,1} f_{2,2}}{3}-\frac{4 f_{1,2} f_{2,1}}{3}+\frac{4 f_{1,3} f_{2,0}}{3}=0; \nonumber\\
&\text{Type}[4,3]:~~-\frac{5 f_{0,2} f_{1,1} f_{3,0}}{2}-\frac{f_{0,2} f_{2,0} f_{2,1}}{3}-\frac{f_{0,3} f_{2,0}^2}{4}+3 f_{0,3} f_{4,0}+\frac{40 f_{1,1}^2 f_{2,1}}{3} \nonumber\\
&~~~~~~~~~~~~~~~-2 f_{1,1} f_{1,2} f_{2,0}-4 f_{1,2} f_{3,1}+3 f_{1,3} f_{3,0}-4 f_{2,1} f_{2,2}=0; \nonumber\\
&\text{Type}[5,2]:~~-\frac{3 f_{0,2} f_{2,0} f_{3,0}}{4}+4 f_{1,1} f_{2,0} f_{2,1}-f_{1,2} f_{2,0}^2+4 f_{1,2} f_{4,0}-8 f_{2,1} f_{3,1}+3 f_{2,2} f_{3,0}=0; \nonumber\\
&\text{Type}[6,2]:~~ \frac{f_{0,2} f_{2,0}^3}{3}-\frac{5 f_{0,2} f_{2,0} f_{4,0}}{3}-\frac{3 f_{0,2} f_{3,0}^2}{8}-\frac{4 f_{1,1}^2 f_{2,0}^2}{3}-\frac{4 f_{1,1}^2 f_{4,0}}{3}+8 f_{1,1} f_{2,0} f_{3,1} \nonumber\\
&~~~~~~~~~~~~~~~+2 f_{1,1} f_{2,1} f_{3,0}-\frac{7 f_{1,2} f_{2,0} f_{3,0}}{2}-2 f_{2,0}^2 f_{2,2}+4 f_{2,0} f_{2,1}^2+4 f_{2,2} f_{4,0}-4 f_{3,1}^2=0; \nonumber\\
&\text{Type}[4,4]_1:~~-\frac{f_{0,2}^2 f_{2,0}^2}{3}-2 f_{0,2}^2 f_{4,0}+6 f_{0,2} f_{1,1} f_{3,1}-3 f_{0,2} f_{1,2} f_{3,0}-2 f_{0,2} f_{2,0} f_{2,2}+2 f_{0,2} f_{2,1}^2 \nonumber\\
&~~~~~~~~~~~~~~~-\frac{9 f_{0,3} f_{1,1} f_{3,0}}{4}-\frac{3 f_{0,3} f_{2,0} f_{2,1}}{2}-f_{0,4} f_{2,0}^2+4 f_{0,4} f_{4,0}-\frac{16 f_{1,1}^4}{3}+4 f_{1,1}^2 f_{2,2} \nonumber\\
&~~~~~~~~~~~~~~~+12 f_{1,1} f_{1,2} f_{2,1}+2 f_{1,1} f_{1,3} f_{2,0}-4 f_{1,3} f_{3,1}=0; \nonumber\\
&\text{Type}[4,4]_2:~~-\frac{f_{0,2}^2 f_{2,0}^2}{18}-\frac{2}{9} f_{0,2} f_{1,1}^2 f_{2,0}+2 f_{0,2} f_{1,1} f_{3,1}+\frac{f_{0,2} f_{1,2} f_{3,0}}{2}-\frac{f_{0,2} f_{2,0} f_{2,2}}{3} \nonumber\\
&~~~~~~~~~~~~~~~~~+\frac{3 f_{0,3} f_{1,1} f_{3,0}}{4}+\frac{f_{0,3} f_{2,0} f_{2,1}}{2} +\frac{16 f_{1,1}^4}{9}-\frac{20 f_{1,1}^2 f_{2,2}}{3}-4 f_{1,1} f_{1,2} f_{2,1}+2 f_{1,1} f_{1,3} f_{2,0} \nonumber\\
&~~~~~~~~~~~~~~~~-4 f_{1,3} f_{3,1}+4 f_{2,2}^2=0; \nonumber\\
&\text{Type}[4,4]_3:~~\frac{4 f_{0,2}^2 f_{2,0}^2}{9}-\frac{5 f_{0,2}^2 f_{4,0}}{3}+\frac{8 f_{0,2} f_{1,1} f_{3,1}}{3}-3 f_{0,2} f_{1,2} f_{3,0}-2 f_{0,2} f_{2,0} f_{2,2}+\frac{4 f_{0,2} f_{2,1}^2}{3}-3 f_{0,3} f_{1,1} f_{3,0} \nonumber\\
&~~~~~~~~~~~~~~~ -3 f_{0,3} f_{2,0} f_{2,1}-\frac{5 f_{0,4} f_{2,0}^2}{3}+4 f_{0,4} f_{4,0}-\frac{64 f_{1,1}^4}{9}+\frac{32 f_{1,1}^2 f_{2,2}}{3}+ \nonumber\\
&~~~~~~~~~~~~~~~16 f_{1,1} f_{1,2} f_{2,1}+\frac{8 f_{1,1} f_{1,3} f_{2,0}}{3}+\frac{4 f_{1,2}^2 f_{2,0}}{3}-4 f_{2,2}^2=0; \nonumber\\
&\text{Type}[5,3]:~~-f_{0,2} f_{1,1} f_{2,0}^2-4 f_{0,2} f_{1,1} f_{4,0}-\frac{f_{0,2} f_{2,0} f_{3,1}}{3}-f_{0,2} f_{2,1} f_{3,0}-\frac{3 f_{0,3} f_{2,0} f_{3,0}}{4} \nonumber\\
&~~~~~~~~~~~~~~~-4 f_{1,1}^3 f_{2,0}+\frac{40 f_{1,1}^2 f_{3,1}}{3}-5 f_{1,1} f_{1,2} f_{3,0}+12 f_{1,1} f_{2,1}^2-f_{1,3} f_{2,0}^2+4 f_{1,3} f_{4,0}-4 f_{2,2} f_{3,1}=0.
\label{conA3}
\end{align}
Here we only write the constraints with label $(a,b),~~a\geq b$, and the other types with label $a<b$ can be derived by exchanging the index of the coefficients appearing in the 
above equations.

\bibliographystyle{JHEP}
\bibliography{PLforRS}

\end{document}